\documentclass[12pt,epsbox]{article}
\usepackage[dvips]{graphicx}
\usepackage{amsmath,amssymb}

\baselineskip=\normalbaselineskip
\renewcommand{\baselinestretch}{1.4}
\makeatletter
\@addtoreset{equation}{section}

\makeatother
\newcommand{\be}{\begin{equation}}
\newcommand{\ee}{\end{equation}}
\newcommand{\ba}{\begin{eqnarray}}
\newcommand{\ea}{\end{eqnarray}}
\newcommand{\nn}{\nonumber \\}

\newcommand{\del}{\partial}
\newcommand{\bra}[1]{\left\langle\,{#1}\,\right|}
\newcommand{\ket}[1]{\left|\,{#1}\,\right\rangle}

\newcommand{\tb}{\widetilde{b}}

\newcommand{\talpha}{\widetilde{\alpha}}

\newcommand{\bc}{\overline{c}}
\newcommand{\br}{\overline{r}}

\newcommand{\bN}{\bar{N}}

\newcommand{\prpr}{\prime\prime}

\begin{document}
\setcounter{page}{0}
\begin{flushright}
\parbox{40mm}{%
RESCEU-32/04 \\
UTAP-497 \\
RIKEN-TH-30 \\
hep-th/0409044 \\
September 2004}
\end{flushright}

\vfill

\begin{center}
{\Large{\bf 
Open String Tachyon in Supergravity Solution
}}
\end{center}

\vfill

\renewcommand{\baselinestretch}{1.0}

\begin{center}
\textsc{Shinpei Kobayashi}
\footnote{E-mail: \texttt{shinpei@utap.phys.s.u-tokyo.ac.jp}},  
\textsc{Tsuguhiko Asakawa}
\footnote{E-mail: \texttt{t.asakawa@riken.jp}} and 
\textsc{So Matsuura}
\footnote{E-mail: \texttt{matsuso@riken.jp}} 

~\\
$^1$ \textsl{Research Center for the Early Universe, \\ 
      The University of Tokyo,  
     7-3-1 Hongo, Bunkyoku, Tokyo 113-0033, JAPAN}  

\vspace{0.5cm}
$^{2, 3}$ \textsl{Theoretical Physics Laboratory, \\ 
      The Institute of Physical and Chemical Research (RIKEN), \\
     2-1 Hirosawa, Wako, Saitama 351-0198, JAPAN } \\ 

\end{center}

\vfill

\begin{center}
{\bf abstract}
\end{center}

\begin{quote}

\small{%
We study the tachyon condensation of the D\={D}-brane system 
with a constant tachyon vev 
in the context of classical solutions of the Type II supergravity. 
We find that the general solution with the symmetry 
$ISO(1,p) \times SO(9-p)$ (the three-parameter solution) 
includes the extremal black $p$-brane solution 
as an appropriate limit of the solution by fixing 
one of the three parameters ($c_1$). 
Furthermore, 
we compare the long distance behavior of the solution 
with the massless modes of the closed strings from 
the boundary state of the D\={D}-brane system 
with a constant tachyon vev.  
We find that we must fix $c_1$ to zero and only two 
parameters are needed  
to express the tachyon condensation of the D\={D}-brane system. 
This means that the parameter $c_1$ does not correspond 
to the tachyon vev of the D\={D}-brane system. 
}
\end{quote}
\vfill

\renewcommand{\baselinestretch}{1.4}

\renewcommand{\thefootnote}{\arabic{footnote}}
\setcounter{footnote}{0}
\addtocounter{page}{1}
\newpage

\section{Introduction}

In recent years, 
unstable D-brane systems have been investigated eagerly 
\cite{Sen,Taylor}. 
Their instability is characterized by the existence of 
a tachyon field on the world-volume  
and they decay to the closed string vacuum or 
the stable BPS states by the tachyon condensation. 
For example, 
if we start from the system with 
a pair of D-brane and \=D-brane,  
it flows to the closed string vacuum \cite{BSFT,AST}. 
Such phenomena are important 
not only for the string theory  
but also for the general relativity.  
In fact, the construction of a dynamical solution 
which represents the tachyon condensation is significant,  
since gravitational or observational objects 
are dynamical in general  
and such dynamical systems will be essential 
to describe realistic objects in terms of the string theory \cite{AKM}. 
 
The aim of this paper is to investigate the unstable D-brane systems 
and their tachyon condensation 
from the viewpoint of 
the classical solutions of the Type II supergravity 
\cite{JM}-\cite{ZZ}. 
In particular, we would like to understand the correspondence between
(microscopic) objects in superstring theory 
and (macroscopic) classical solutions in the supergravity.
It is known that there is such a correspondence between 
the BPS D-branes and the extremal black $p$-brane solutions 
\cite{HS}. 
However, for unstable D-brane systems, it is not fully understood
what their classical counterparts are, although there are some
discussions \cite{BMO,LR1}. 

Previously, in \cite{DiVecchia}, it is argued that there is no 
such correspondence for unstable D-brane systems. 
The main point of their discussion is as follows: 
the supergravity 
approximation is valid only when curvature effects are small, 
and it can be realized by 
sufficiently large number of coincident D-branes. 
However, the unstable D-brane systems
do not satisfy the no-force condition, unlike BPS cases, 
then, we cannot consider multiple coincident D-branes. 
As opposed to this observation, we argue here that 
it is still possible to study the correspondence. 
As discussed in \cite{DiVecchia,DiVecchia2}, 
once we fix a source of closed strings, 
we can determine the solution of the equation of motion 
of the closed string field theory uniquely. 
Thus, the problem is whether we can construct a source 
corresponding to an unstable D-brane system 
where the supergravity approximation is valid. 
Here, the important fact is that we can express any type of 
D-branes as boundary states with appropriate boundary 
interactions \cite{BSFT,AST}. 
In particular, it is possible to construct many number of coincident 
unstable D-branes in this formulation, 
and this argument is irrespective of  
whether the D-brane system is stable or not. 
Since the validity of the supergravity approximation is the 
same as in the BPS case, the comparison is meaningful as long as 
sufficiently large number of coincident unstable D-branes.
Note that if the source is static, then the solution is also static.
From this observation, it is possible 
to deal with an unstable D-brane system in the context of 
a classical solution of the supergravity.

As a remark we comment on the no-force condition. 
The no-force condition is related to dynamical issues, 
that is, it is the condition that the interaction between 
several D-branes vanishes, 
which guarantees that the superposition of 
several D-branes also becomes a consistent source of 
bulk closed strings even if we take into account 
the interaction between D-branes. 
Actually, 
an unstable D-brane system does not satisfy the no-force condition 
and is expected to 
receive a backreaction effect and decay to some stable one,
if we take into account such interactions.
The full solution is obtained by solving the equation of motion of gravity
coupled with a dynamical source (matter) and the matter equation of motion
simultaneously\cite{AKM}. 
The classical solution we will consider in this paper is thought to be 
the zero-th order approximation to the full solution. 
There are also other issues, such as $\alpha'$ and loop correction, 
but we will not discuss them here.

In this paper, we consider a system of 
coincident $N$ D$p$-branes and $\bN$ \=D$p$-branes 
(for simplicity we restrict ourselves to $N>\bN$)  
with an excitation of an open string tachyon on them. 
This system is a typical example of the unstable D-brane systems. 
For simplicity, 
we assume that the tachyon profile on the world-volume 
is constant. 
In this setting, the system possesses 
the global symmetry $ISO(1,p)\times SO(9-p)$. 
Although any number of D\=D-pairs can 
be annihilated by the tachyon condensation in general,  
we concentrate on the case where the final state is 
BPS saturated, i.e. the system of $(N-\bN)$ D-branes.%
\footnote{ 
If we assume $(N - \bN) \gg 1$, 
the supergravity description is valid. }
On the other hand, 
the general solution with the same symmetry $ISO(1,p)\times SO(9-p)$ 
in the Type II supergravity 
is known as the ``three-parameter solution'' \cite{ZZ}. 
Therefore, it is natural that the three parameters of the solution 
have a relationship to physical parameters of the D\={D}-brane system. 
In fact, in \cite{BMO,LR1}, 
it is argued that the three parameters 
correspond to microscopic quantities, that is, 
the number of D$p$-branes $N$,  
the number of \={D$p$}-branes $\bar{N}$ 
and the tachyon vev $\langle T \rangle$, respectively. 
In this paper, we re-examine these correspondences in two ways. 
First, we introduce a new parametrization for the 
three-parameter solution and show that the solution
becomes the extremal $p$-brane solution \cite{HS} by tuning one of the
new parameter $(\epsilon)$ to zero while fixing another parameter $(c_1)$
to an arbitrary value. 
It means that $c_1$ cannot correspond to the tachyon vev
$\langle T \rangle$, as opposed to the proposal 
which has been made so far \cite{BMO,LR1}. 
Next, 
we directly compare the three-parameter
solution with a boundary state which expresses the non-BPS D-brane
system in considering, in order to obtain more quantitative 
correspondence between microscopic parameters 
and macroscopic ones.  
In this investigation, we use the technique given in \cite{DiVecchia}. 
As we mentioned before, we can apply this technique even to non-BPS 
boundary states 
although it is originally used to show the correspondence between the
extremal black $p$-brane solution and the BPS boundary state. 
The main advantage of this method, 
compared to the previous approach \cite{BMO,LR1},  
is the usage of the off-shell boundary state for the D\={D}-brane system
with a tachyon vev, which satisfies Sen's conjecture. 
As a result, we will find that we need only two parameters 
to express the D\=D-brane system. 
The physical meaning of this parameter is an open question at present,
but it will be discussed in \cite{AKM2}. 
(For recent works, see \cite{LR2}.) 

The organization of this paper is as follows. 
In the next section, we review the ten-dimensional 
three-parameter solution. 
In the section 3, 
which is the main part of this paper, 
we show that the three-parameter solution becomes 
the extremal black $p$-brane solution by taking 
an appropriate limit with one of the parameters 
of the solution fixing to an arbitrary value. 
We investigate the long distance behavior of 
the three-parameter solution 
and compare it with the boundary state corresponding to 
the system of $N$ D-branes and $\bN$ \=D-branes 
with a constant tachyon vev. 
We determine the values of the three parameters of the solution 
so that it represents the tachyon condensation 
of the D\=D-brane system.  
Section 4 is devoted to the conclusion and the discussion.

\section{Review of the Three-parameter Solution in Ten-dimensions}

In this section, we review the three-parameter solution 
given in \cite{ZZ}.%
\footnote{
In \cite{ZZ}, a general solution 
with the symmetry $ISO(p)\times SO(D-p-1)$ is constructed 
as the ``four-parameter solution''. }  
Since we are interested in the string theoretical interpretation 
of the solution, we only consider the ten-dimensional 
Type II supergravity. For the three-parameter solution for an 
arbitrary dimensionality, see the appendix A.  

We start with the ten-dimensional action, 
\begin{equation}
S = \frac{1}{2\kappa^2} \int d^{10} x \sqrt{-g} 
    \left[ 
    R -\frac{1}{2}(\del \phi)^2 
    -\frac{1}{2(p+2)!} e^{\frac{3-p}{2} \phi} |F_{p+2}|^2 
    \right], 
\label{ten-3para-action}
\end{equation}
where $F_{(p+2)}$ denotes the $(p+2)$-form field strength which 
relates to the $(p+1)$-form potential of the RR-field $\mathcal{A}^{(p+1)}$ 
as $F_{(p+2)} = d \mathcal{A}_{(p+1)}$. 
Since we consider the solution with 
the symmetry $ISO(1,p)\times SO(9-p)$, 
we impose the ansatz, 
\begin{align}
ds^2 &= g_{MN} dx^M dx^N \nonumber \\ 
     &= e^{2A(r)} \eta_{\mu\nu}dx^{\mu}dx^{\nu} 
      + e^{2B(r)} \delta_{ij} dx^i dx^j \nonumber \\
     &=e^{2A(r)} \eta_{\mu\nu}dx^{\mu}dx^{\nu} 
      + e^{2B(r)} (dr^2 + r^2 d\Omega_{(8-p)}^2 ),  \nonumber \\
\phi &= \phi(r), \nonumber \\ 
\mathcal{A}^{(p+1)} &= e^{\Lambda(r)}\ 
             dx^0 \wedge dx^1 \wedge \cdots \wedge dx^p,  
\label{ten-3para-ansatz}
\end{align}
where $\mu,\nu=0,\cdots,p$ are indices of the longitudinal  
directions of the $p$-brane, 
$i,j=p+1,\cdots,9$ express the orthogonal directions, 
and $M=(\mu,i)$. 
Under this ansatz, the equations of motion become
\begin{align}
A^{\prpr} + \left( (p+1)A^{\prime} 
+(7-p)B^{\prime} 
+\frac{8-p}{r}
\right)A^{\prime} 
&= \frac{7-p}{16} S^2, \nonumber \\
B^{\prpr} + \left( (p+1)A^{\prime} 
+(7-p)B^{\prime} 
+\frac{15-2p}{r} 
\right)B^{\prime} + \frac{p+1}{r}A^{\prime} 
&= -\frac{p+1}{16} S^2, \nonumber \\ 
(p+1)A^{\prpr} + (8-p)B^{\prpr} 
+(p+1)A^{\prime 2}   \nonumber 
-(p+1)A^{\prime}B^{\prime}
+\frac{8-p}{r}B^{\prime} 
+\frac{1}{2} \phi^{\prime 2} 
&= \frac{7-p}{16} S^2, \nonumber \\ 
\phi^{\prpr} + \left(
                   (p+1)A^{\prime} 
                   +(7-p)B^{\prime} 
                   +\frac{8-p}{r}
                  \right) \phi^{\prime} 
&= -\frac{3-p}{4} S^2, \nonumber \\
 \left( 
   \Lambda^{\prime} 
    e^{\Lambda + \frac{3-p}{2}\phi -(p+1)A+(7-p)B} r^{8-p}
 \right)^{\prime} &= 0, 
\label{eom}
\end{align}
where 
\begin{equation}
S = \Lambda^{\prime} e^{\Lambda+\frac{3-p}{4}\phi-(p+1)A},  
\end{equation}
and the prime denotes the derivative with respect to $r$. 

The authors in \cite{ZZ} show that the general 
asymptotically flat solution 
of the equations (\ref{eom}) is given by 
\begin{align}
\label{sln-1}
A(r) &= \frac{(7-p)(3-p)c_1}{64} h(r) 
        -\frac{7-p}{16} 
        \ln \left[
            \cosh (k h(r)) -c_2 \sinh (k h(r))
            \right], \\
\label{sln-2}
B(r) &= \frac{1}{7-p} \ln [f_-(r) f_+(r)] \nonumber \\
        & \hspace{0.5cm}-\frac{(p+1)(3-p) c_1}{64} h(r) 
        +\frac{p+1}{16} 
        \ln \left[
            \cosh (k h(r)) -c_2 \sinh (k h(r))
            \right], \\
\label{sln-3}
\phi(r) &= \frac{(p+1)(7-p) c_1}{16} h(r) 
           +\frac{3-p}{4} 
            \ln \left[
            \cosh (k h(r)) -c_2 \sinh (k h(r))
            \right], \\
\label{sln-4}
e^{\Lambda(r)} &= - \eta  (c_2^2 -1)^{1/2}\  
                  \frac{\sinh(k h(r))}
                   {\cosh (k h(r)) -c_2 \sinh (k h(r))},
\end{align}
where 
\begin{align}
f_{\pm}(r) &\equiv 1 \pm \frac{r_0^{7-p}}{r^{7-p}}, \\
h(r) &\equiv \ln \left( \frac{f_-}{f_+} \right), \\
 k &\equiv \pm
 \sqrt{ \frac{2(8-p)}{7-p} - \frac{(p+1)(7-p)}{16} c_1^2 } 
  \nonumber \\
 &\equiv \pm \frac{\sqrt{(p+1)(7-p)}}{4}\, \sqrt{c_m^2-c_1^2}, 
 \qquad \left(c_m = \sqrt{\frac{32(8-p)}{(p+1)(7-p)^2} } \right)
 \\
\eta &= \pm 1.
\end{align}
The three parameters,%
\footnote{
We have labeled $c_3$ of \cite{ZZ} as $c_2$ and $k$ as $-k$ 
according to \cite{BMO}.}
$r_0,\ c_1,\ c_2$, are the integration constants 
that parametrize the solution. 
As discussed in \cite{BMO}, the region of the parameters 
in the solution (\ref{sln-1})--(\ref{sln-4}) is 
\begin{align}
 c_1 &\in (0, c_m), \\
 c_2 &\in (-\infty,-1)\cup(1,\infty), \\
 r_0^{7-p} &\in {\mathbf R}, 
\end{align}
where we have already fixed the ${\mathbf Z}_2$ symmetries 
of the solution, 
\begin{align}
 (r_0^{7-p},c_1,c_2,sgn(k),\eta) &\to 
 (r_0^{7-p},c_1,-c_2,-sgn(k),-\eta),  \nonumber \\
 (r_0^{7-p},c_1,c_2,sgn(k),\eta) &\to 
 (-r_0^{7-p},-c_1,c_2,-sgn(k),\eta), 
\end{align}
by choosing $c_1 \ge 0$ and $k\ge 0$. 
Furthermore, we have a degree of freedom to choose 
the signs of $r_0^{7-p}$ and $c_2$. 
In the next section, we will discuss 
the extremal black $p$-brane limit of the solution. 
To take this limit consistently, we must choose the branch,  
\begin{equation}
 (r_0^{7-p} \ge 0, c_2 \ge 0),\quad {\rm or}\quad 
 (r_0^{7-p} \le 0, c_2 \le 0).
\end{equation}
For simplicity, we choose $(r_0^{7-p} \ge 0, c_2 \ge 0)$ 
in this paper.%
\footnote{
In \cite{BMO}, the authors choose 
$(r_0^{7-p} \le 0, c_2 \le 0)$ for $p=0,\cdots,3$ and 
$(r_0^{7-p} \ge 0, c_2 \ge 0)$ for $p=3,\cdots,6$ 
by imposing that $M$ should decrease monotonically 
as $c_1$ increases, since they identify $c_1$ as the tachyon vev. 
In this paper, however, we conclude that $c_1$ is not 
the vev of the open string tachyon, and thus, 
there is still an ambiguity in the signs of $r_0^{7-p}$ and $c_2$. 
}

From the view point 
of the gravity theory, 
the three-parameter solution describes a 
charged dilatonic black object. 
Thus, the RR-charge $Q$ and the ADM mass $M$ 
are natural quantities to characterize the solution.%
\footnote{
In addition to $Q$ and $M$, it is often useful to define 
the ``dilaton charge'' $D$.  
In the case of the three-parameter solution \cite{OY}, 
\begin{equation}
D = N_p \left[16c_2 k -4(p+1)c_1 \right] r_0^{7-p}.  \nonumber 
\end{equation}
Instead of $(c_1,c_2,r_0)$, we can use $(Q,M,D)$ to parametrize 
the solution. 
} 
For convenience, we consider wrapping the spatial world-volume 
directions on a torus $T^p$ of volume $V_p$. 
The RR-charge is given by an appropriate surface integral 
over the sphere-at-infinity in the transverse directions 
\cite{BMO,stelle};  
\begin{equation}
Q = 2\eta N_p (c_2^2- 1)^{1/2} k r_0^{7-p}, 
\label{RR-charge}
\end{equation} 
where 
\begin{equation}
N_p \equiv \frac{(8-p)(7-p) \omega_{8-p}V_p}{16\kappa^2},  
\end{equation}
and $\omega_d=\frac{2\pi^{(d+1)/2}}{\Gamma((d+1)/2)}$ 
is the volume 
of the unit sphere $S^d$. 
The ADM mass is defined as \cite{Mal,MP} 
\begin{equation}
g_{00} = -1 + \frac{2\tilde{\kappa}^2 M}{(8-p)\omega_{8-p}r^{7-p}} 
            +\mathcal{O}\left( \frac{1}{r^{2(7-p)}}\right), 
\end{equation}
where the metric is written in Einstein frame and 
$\tilde{\kappa}^2 \equiv \kappa^2/V_p$. 
Using this definition, we see that the ADM mass of 
the three-parameter solution is \cite{BMO} 
\begin{equation}
M = N_p \left(\frac{3-p}{2}c_1 + 2c_2 k \right) r_0^{7-p}. 
\label{ADM-mass}
\end{equation}

\section{D\=D-brane System in the Three-parameter Solution}

In the previous section, we reviewed the ten-dimensional 
three-parameter solution and defined the physical quantities 
$Q$ and $M$. 
In this section, 
we first show that we can obtain the extremal black $p$-brane 
solution as an appropriate limit of the three-parameter solution 
for arbitrary $c_1$ by using a reparametrization. 
Then we introduce another convenient parametrization 
for dealing with the tachyon condensation, 
in the sense that we can change the solution 
without changing the value of the RR-charge. 
By comparing the behavior of the solution 
with the boundary state corresponding to 
the system of $N$ D-branes and $\bN$ \=D-branes 
with a constant tachyon vev, 
we show that we must choose $c_1=0$ to express 
the tachyon condensation of the D\=D-brane system 
by the three-parameter solution.

\subsection{Black $p$-brane solution in the three-parameter solution 
and new parametrizations}

Let us first show that the three-parameter solution coincides 
with the extremal black $p$-brane solution under a certain limit 
for arbitrary $c_1$.%
\footnote{
After almost completing this work, we found that the same idea 
has been developed independently by the authors of \cite{LR2}.  
}
The ten-dimensional extremal black $p$-brane solution ($p<7$) is 
given by \cite{HS}  
\begin{align}
ds^2 &= f_p(r)^{-\frac{7-p}{8}} \eta_{\mu\nu} dx^{\mu}dx^{\nu} 
       +f_p(r)^{\frac{p+1}{8}} \delta_{ij} dx^{i}dx^{j}, 
\label{extreme-1}
\\
e^{\phi}(r) &= f_p(r)^{\frac{3-p}{4}}, 
\label{extreme-2}
\\
e^{\Lambda(r)} &= -\eta (f_p(r)^{-1} -1),
\label{extreme-3}
\end{align}
where $f_p(r)$ is defined as 
\begin{equation}
f_p(r) = 1 +\frac{\mu_0}{r^{7-p}},  
\label{f_p}
\end{equation}
and $\mu_0$ is the only parameter of 
this solution that 
is proportional to the RR-charge 
of the D$p$-branes. 
To realize the solution (\ref{extreme-1})--(\ref{extreme-3}) 
in the three-parameter solution, 
we introduce a parametrization by $\br_0^{7-p}$ and $\epsilon$ as
\begin{align}
r_0^{7-p} = \epsilon \  \br_0^{7-p}, \qquad
c_2 = \frac{1}{\epsilon}. \qquad (0\le \epsilon \le 1)
\label{epsilon'}
\end{align}
When $\epsilon \to 0$, the three-parameter solution coincides with 
the extremal black $p$-brane solution (\ref{extreme-1})-(\ref{extreme-3}) 
with the identification,%
\footnote{
In \cite{BMO}, the black $p$-brane solution is given as the 
$\epsilon\to 0$ limit of 
\begin{align}
r_0^{7-p} = \epsilon^{1/2} \br_0^{7-p}\ , \qquad
k = \epsilon^{1/2} \bar{k}\ , \qquad
c_2 = \frac{\bc_2}{\epsilon}. \nonumber
\end{align}
Since $k \to 0$ means $c_1 \to c_m$, 
this is a particular example of (\ref{epsilon'}). 
} 
\begin{equation}
\mu_0 = 2k \br_0^{7-p}. 
\label{identify}
\end{equation}
In the parametrization (\ref{epsilon'}), 
the RR-charge (\ref{RR-charge}) 
and the ADM mass (\ref{ADM-mass}) 
are written as 
\begin{align}
\label{Q in epsilon'}
 Q &= 2\eta N_p (1-\epsilon^2)^{1/2} k \br_0^{7-p}, \\ 
 M &= 2 N_p (1+\frac{3-p}{4k}c_1\epsilon)  k \br_0^{7-p}. 
\end{align} 
If we set $c_1=0$,%
\footnote{
This is the case of the D\=D-brane system 
we consider in this article. 
See the next subsection. 
}
$M$ is invariant under the change of $\epsilon$, 
and $Q$ and $M$ satisfy the relation, 
\begin{equation}
 M^2 - Q^2 = ( 2 N_p k {\br_0^{7-p}} \epsilon)^2, 
\end{equation} 
thus 
the parameter $\epsilon$ can be regarded as 
a ``non-extremality parameter''. 
Note that, 
in the $\epsilon\to 0$ limit, 
the RR-charge and the ADM mass become 
\begin{equation}
 \left|Q\right| = M = 2N_p k \br_0^{7-p}, 
\end{equation}
for arbitrary $c_1$. 
Thus we conclude that $c_1$ has nothing to do with the tachyon vev, 
since the solution becomes BPS for arbitrary $c_1$.

The above parametrization, however, is not suitable 
for our purpose to analyze the tachyon condensation 
of the D\=D-brane system, 
since the RR-charge depends on $\epsilon$. 
As is explained in detail in the next subsection, 
not the RR-charge but the mass of the D\={D}-brane system 
will change as the tachyon condensates. 
Thus, we define another set of parameters $(v,\mu_0)$ 
to fix $Q$ as 
\begin{equation}
 r_0^{7-p} \equiv \frac{v\mu_0}{2k},\qquad 
 c_2^2-1 \equiv \frac{1}{v^2}. \qquad (0\le v \le \infty)
 \label{new-para}
\end{equation} 
It is easy to show that the $v\to 0$ limit of the 
solution is again the extremal black $p$-brane solution 
(\ref{extreme-1})--(\ref{extreme-3}). 
In this parametrization, 
the RR-charge and the ADM mass are expressed as 
\begin{align}
 Q &= \eta N_p \mu_0, \\
 M &= N_p \left(\sqrt{1+{v^2}}
                  +\frac{3-p}{4k}c_1 v \right)\mu_0, 
\label{new-QM}
\end{align}
in which the RR-charge does not depend on 
$v$ and $c_1$ as announced. 
Note that $v$ works as the ``non-extremality parameter'' 
as $\epsilon$ does, but $v$ denotes how much the ADM 
mass exceeds the RR-charge.     

For the later discussion, 
we write down the long distance behavior of this solution;  
up to the order $1/r^{7-p}$. 
The result is
\begin{align}
\label{asymp-1}
 e^{2A(r)} &= 1
  -\frac{7-p}{8}
   \left(\sqrt{1+v^2}+\frac{3-p}{4} \frac{c_1 v}{k} \right)
   \frac{\mu_0}{r^{7-p}}
  + {\cal O}\left(1/r^{2(7-p)}\right),  \\
\label{asymp-2}
 e^{2B(r)} &= 1
  +\frac{p+1}{8}
    \left(\sqrt{1+v^2}+\frac{3-p}{4} \frac{c_1 v}{k} \right)
   \frac{\mu_0}{r^{7-p}}
  + {\cal O}\left(1/r^{2(7-p)}\right),  \\
\label{asymp-3} 
\phi(r) &= \left(
  \frac{3-p}{4}\sqrt{1+v^2} - \frac{(p+1)(7-p)}{16}\frac{c_1 v}{k} 
 \right)
    \frac{\mu_0}{r^{7-p}}
  + {\cal O}\left(1/r^{2(7-p)}\right),  \\
 e^{\Lambda(r)} &= \eta \frac{\mu_0}{r^{7-p}}
  + {\cal O}\left(1/r^{2(7-p)}\right). 
\label{asymp-4}
\end{align}

\subsection{Comparison with the boundary state}

In this subsection, we show that the D\={D}-brane system with 
a constant tachyon vev corresponds to 
the three-parameter solution of $c_1=0$. 
This means that we need only two 
parameters to express the D\={D}-brane system in 
the low-energy gravity theory. 
To show it, we use the technique to reveal the correspondence between 
the extremal black $p$-brane solution 
and the boundary state established in \cite{DiVecchia}. 

Let us first briefly review the system of $N$ D$p$-branes 
and $\bar{N}$ \=D$p$-branes with a constant tachyon vev. 
The gauge symmetry of the low energy effective theory 
on the world-volume is $U(N)\times U(\bN)$. 
There is a complex tachyon field $T(x)$ on it that is in the 
bi-fundamental representation of the gauge group. 
In this paper, we consider the case where the $\bN$ D\=D-pairs vanish
and $(N-\bN)$ D$p$-branes remain. 
Namely, we decompose the $\bN \times N$ matrix by $\bN \times (N-\bN)$ and 
$\bN \times \bN$ components and set the tachyon profile as
\begin{eqnarray}
T(x)=\left(
\begin{array}{c}
0\\
\hline
T
\end{array}
\right),
\end{eqnarray}
where $T$ is a constant $\bN \times \bN$ matrix.
Note that, since the other open string excitations, 
gauge fields, scalar fields and a non-constant tachyon  
break the global symmetry $ISO(1,p)\times SO(9-p)$, 
we do not consider them here.
It is known that this system can be expressed by the boundary state 
on which the boundary interaction for the tachyon field 
is turned \cite{BSFT,AST}. 
For our case, the NSNS part of the boundary state 
is then given by 
\begin{align}
\ket{B_p;T}_{NSNS} 
 &= \frac{T_p}{2}
 \left[
 (N-\bN) + 2{\rm tr\,} e^{-|T|^2}
 \right] 
 \delta^{(9-p)}(x^i) \nonumber \\ 
 & \hspace{0.5cm}\times  
 \exp \left[
 -\sum_{n=1}^{\infty} \frac{1}{n}
 \alpha_{-n}^{M}S_{MN} \talpha_{-n}^{N}
 \right]
 \sin \left[
 -\sum_{r=1/2}^{\infty} 
 b_{-r}^{M} S_{MN} \tb_{-r}^{N}
 \right] \ket{0}, 
\label{boundary state}
\end{align}
where $S_{MN} = (\eta_{\mu\nu}, -\delta_{ij})$, 
$\ket{0}$ is the Fock vacuum of a closed string in the NSNS sector, 
and 
$\alpha_{-n}^M$ and $b_{-r}^M$ are the creation operators 
of the modes of the world-sheet bosons and fermions, respectively. 
The trace is taken over $\bN \times \bN$ matrices.
Note that only the difference from the BPS boundary state is
the effect of the constant tachyon vev, which contributes 
only to the overall numerical factor 
$T_p [(N-\bN) + 2{\rm tr\,} e^{-|T|^2}]$ in (\ref{boundary state}).
The first term of this factor corresponds to the tension of 
the $(N-\bN)$ D-branes, 
and the second term is 
the tachyon potential $V(T)$ for the $\bN$ D\=D-pairs,  
which satisfies Sen's conjecture $V(T=0)-V(T_{min})=2\bN T_p$.
Thus, when $T=0$ (i.e. ${\rm tr\,} e^{-|T|^2}=\bN$), 
the factor is the tension of 
the sum of $N$ D-branes and $\bN$ \=D-branes, 
while, when $|T|\sim\infty$, 
the tension becomes that of $(N-\bN)$ D-branes. 
On the other hand, 
the boundary state of the RR-sector is exactly the same as 
that of the BPS D$p$-brane 
under the excitation of the constant tachyon $T$,
since the RR $(p+1)$-form charge conserves 
and other kinds of charges do not appear. 

According to the discussion in \cite{DiVecchia}, 
the long distance behavior of the classical solution of 
the Type II supergravity corresponding to the D\=D-brane system 
can be read off from the boundary state (\ref{boundary state}). 
Following \cite{DiVecchia}, 
we first define the quantity, 
\begin{align}
J^{MN}(k) &\equiv \bra{0;k} 
              b_{1/2}^{M}\tb_{1/2}^{N} D \ket{{B}_p;T}_{NSNS} \nonumber\\
          &= -\frac{T_p}{2} 
             \left[
              (N-\bN) + 2{\rm tr\,} e^{-|T|^2}
             \right]
               \frac{V_{p+1}}{k_i^2} S^{MN}.
\end{align} 
Here $k_i$ is the momentum of the transverse direction to the 
brane. 
We can evaluate 
the long distance behavior of the graviton and the dilaton 
by multiplying the corresponding polarization tensors,  
\begin{align}
\epsilon_{MN}^{(h)} &= \epsilon_{MN}^{(h)}, 
\hspace{1cm} 
\epsilon_{MN}^{(h)} \eta^{MN} = \epsilon_{MN}^{(h)} k^M = 0, \\
\epsilon_{MN}^{(\phi)} 
 &= \frac{1}{2\sqrt{2}}[\eta_{MN}- k_M l_N -k_N l_M], 
\hspace{1cm} 
k\cdot l =1,\ k^2 = l^2 =0, 
\end{align} 
to $J_{MN}(k)$; 
\begin{align}
\label{graviton-k}
\widetilde{h}_{MN}^{(1)}(k) &= 
 2\kappa \times \left(
  J_{MN}(k)
   -\frac{J^{MN}(k) \epsilon_{MN}^{(\phi)}}
         {\eta^{MN} \epsilon_{MN}^{(\phi)} } \eta_{MN} 
 \right) \nn
               &= \left[
                     (N-\bN) + 2{\rm tr\,} e^{-|T|^2}
                    \right]
                   \frac{2\kappa T_p V_{p+1}}{k_i^2} 
                  \left(
                    -\frac{7-p}{8} \eta_{\mu\nu},\ 
                  \frac{p+1}{8} \delta_{ij}
                  \right),  \\
\widetilde{\phi}^{(1)}(k) &= \sqrt{2}\kappa \times 
J^{MN}(k) \epsilon_{MN}^{(\phi)} \nn 
                &= \left[
                    (N-\bN) + 2{\rm tr\,} e^{-|T|^2}
                   \right] 
                    \frac{2\kappa T_p V_{p+1}}{k_i^2} 
                   \cdot 
                   \frac{3-p}{4},  
\label{dilaton-k}
\end{align}
where the additional factors, $2\kappa$ and $\sqrt{2}\kappa$,
are necessary to compare with 
the classical solution. (For detail, see \cite{DiVecchia}.)
For the RR $(p+1)$-form field, the similar calculation leads to 
the result, 
\begin{align}
 e^{\widetilde{\Lambda}^{(1)}}(k) 
= \eta (N-\bar{N})\frac{2\kappa T_p V_{p+1}}{k_i^2}. 
\label{ramond-k}
\end{align} 
To compare the result from the boundary state 
(\ref{graviton-k})--(\ref{ramond-k}) with the three-parameter solution, 
we use the Fourier transformation, 
\begin{equation}
\int d^{10}x e^{i k_i \cdot x^i} \frac{1}{(7-p)\omega_{8-p}} 
                                 \frac{1}{r^{7-p}}
= \frac{V_{p+1}}{k_i^2}, 
\end{equation}
then we rewrite 
(\ref{graviton-k})--(\ref{ramond-k}) as functions of $r$; 
\begin{align}
 \label{graviton-s}
 h_{MN}^{(1)}(r)&=
  \left[ (N-\bN) + 2{\rm tr\,} e^{-|T|^2} \right]
 \frac{2\kappa T_p}{(7-p)\omega_{8-p}}\frac{1}{r^{7-p}} 
 \left(
 -\frac{7-p}{8} \eta_{\mu\nu},\ 
 \frac{p+1}{8} \delta_{ij}
 \right),  \\
 \label{dilaton-s}
 \phi^{(1)}(r) &= 
 \left[ (N-\bN) + 2{\rm tr\,} e^{-|T|^2} \right] 
 \frac{2\kappa T_p}{(7-p)\omega_{8-p}}\frac{1}{r^{7-p}}
 \cdot 
 \frac{3-p}{4},  \\ 
  e^{{\Lambda}^{(1)}}(r) &= 
 \eta (N-\bar{N})\frac{2\kappa T_p}{(7-p)\omega_{8-p}}\frac{1}{r^{7-p}}.
\label{ramond-s}
\end{align}

Let us compare the long distance behavior 
of the classical solution (\ref{asymp-1})--(\ref{asymp-4}) 
with the result from the boundary state 
(\ref{graviton-s})--(\ref{ramond-s}). 
First, from the results of the RR field (\ref{asymp-4}) 
and (\ref{ramond-s}), we can identify $\mu_0$ as 
\begin{equation}
 \mu_0 = \frac{2\kappa T_p (N-\bN)}{(7-p)\omega_{8-p}}\,.  
\end{equation}
Next, from the graviton and the dilaton, 
we see that the parameters must satisfy the following relations
simultaneously;
\begin{equation}
 1+\frac{2}{N-\bar{N}}{\rm tr\,}e^{-|T|^2} 
  = \sqrt{1+v^2}+\frac{3-p}{4}\frac{c_1 v}{k}
   = \sqrt{1+v^2}-\frac{(p+1)(7-p)}{4(3-p)}\frac{c_1 v}{k}, 
\end{equation}
which are consistent only when $c_1=0$. 
From this consideration, we conclude that 
the three-parameter solution with 
\begin{align}
 c_1&=0, \nn
 \label{final}
 \mu_0 &= \frac{2\kappa T_p (N-\bN)}{(7-p)\omega_{8-p}}, \\
 \sqrt{1+v^2} &= 1+\frac{2}{N-\bar{N}}{\rm tr\,}e^{-|T|^2}, \nonumber
\end{align} 
corresponds to the system of $N$ D-branes and $\bN$ \=D-branes 
with a constant tachyon vev. 
We emphasize
that the vev of the open string tachyon between 
the D-branes and the \=D-branes does not correspond to $c_1$, 
but corresponds to $v$, as opposed to the previous expectation. 

Once this correspondence is obtained, we can easily relate 
the ADM mass $M$ and the RR-charge $Q$ 
to the quantities of the D\=D-brane system as 
\begin{align}
Q &= \eta T_p \left(N-\bN \right)\frac{(8-p)V_p}{8\kappa}, \\ 
M &= T_p \left[ \left(N-\bN \right) +2{\rm tr\,} e^{-|T|^2} \right]
 \frac{(8-p)V_p}{8\kappa}. 
\end{align} 
This is quite natural result that the overall coefficient of 
the NSNS and the RR 
boundary state is directly related to the mass 
and the RR-charge, respectively, 
as in the BPS cases.

\section{Conclusion and Discussion}

In this paper, we discussed the tachyon condensation of 
the system of $N$ D-branes and $\bar{N}$ \=D-branes 
with a constant tachyon vev $T$ using the general solution 
of the Type II supergravity with the symmetry 
$ISO(1,p)\times SO(9-p)$. 
Since the solution is parametrized by the three parameters 
$c_1,c_2$ and $r_0$, 
it is called as the ``three-parameter solution''. 
Introducing the new parameters $\bar{r}_0$ and $\epsilon$ 
instead of $c_2$ and $r_0$, 
we found that the three-parameter solution becomes 
the extremal black $p$-brane solution by taking 
$\epsilon \to 0$ for arbitrary $c_1$. 
This means that $c_1$ is not  related to tachyon vev.

In order to relate the three-parameter solution with 
the D\=D-brane system more directly, 
we compared the long distance 
behavior of the three-parameter 
solution with the massless modes of the closed string 
from the D\=D-brane boundary state. 
We can apply the relation between the classical 
solution and the boundary state to our case, 
although the D\=D-brane system is a non-BPS state 
except at the end of the tachyon condensation. 
In this connection, we can similarly apply this method to 
any other sources that are non-BPS or unstable. 
  
When we compared the solution with the boundary state, 
we introduced the parametrization  
$\{\mu_0, v\}$ so that we can fix the RR-charge. 
In fact, the RR-charge is conserved during the tachyon condensation, 
and neither the parametrization $\{c_2, r_0\}$ nor the parametrization
$\{\epsilon, \bar{r}_0 \}$ are suitable to such a situation. 
In contrast, the parametrization $\{ \mu_0, v \}$ is convenient 
for our purpose because it makes the RR-charge depend only on 
$\mu_0$, and we can change the ADM mass without changing $\mu_0$.

By comparing the long distance behavior of the three-parameter 
solution with the massless modes of the closed string from the boundary 
state of the D\=D-brane system, 
we found that we must fix $c_1$ to zero 
and only two parameters $\mu_0$ and $v$ (or  $c_2$ and $r_0$) 
are needed to express the D\=D-brane system. 
This result means that  
we can see the tachyon vev not as an independent parameter in the 
supergravity solution 
but as a part of the ADM mass. 
This fact is consistent with the form of the boundary state, 
i.e. the tachyon vev is included in the coefficient 
of the NSNS sector of the boundary state as shown in 
(\ref{boundary state}), and the coefficient is nothing but the 
ADM mass of the solution. 
So we concluded that the D\=D-brane system corresponds to 
the three-parameter solution of $c_1=0$. 
If $c_1$ turned on, the long distance behavior of the 
three-parameter solution can not be reproduced by the boundary state 
of the D\=D-brane system.  
This means that $c_1$ relates to another physical quantity 
in the string theory, which will be discussed in the forthcoming 
paper \cite{AKM2}.

At the end of this paper, we would like to note that the 
parameter $c_1$ is not so unfamiliar at least in the general 
relativity. 
In fact, $c_1$ has something to do with the quantity which has 
been known as a ``scalar charge'' in the general relativity. 
To see this explicitly, let us see the Janis-Newman-Winicour solution 
or the Wyman solution \cite{JNW}-\cite{Virb}. It is the 
four-dimensional and SO(3) symmetric solution of the Einstein 
equation with a free scalar field $\phi$. 
The concrete form is  
\begin{align}
ds^2 &= -\left[ \frac{f_-(r)}{f_+(r)} \right]^{2\gamma} dt^2 
        +\left[f_-(r)\right]^{2-2\gamma} 
         \left[f_+(r)\right]^{2+2\gamma}
          \left(dr^2 + r^2 d\Omega_{(2)}^2 \right), \\
\phi(r) &= \sqrt{\frac{4q^2}{m^2+q^2}} 
           \ \ln \left[ \frac{f_-(r)}{f_+(r)} \right], 
\label{scalar charge}
\end{align}
where $f_{\pm}(r)$ is defined as 
\begin{equation}
f_{\pm}(r) = 1 \pm \frac{\sqrt{m^2+q^2}}{r}, \hspace{0.5cm} 
\gamma = \sqrt{\frac{m^2}{m^2 + q^2}}. 
\end{equation}
Here $q$ is the so-called scalar charge, 
which relates to the scalar field $\phi$ as shown 
in (\ref{scalar charge}).
If we set $q=0$, the scalar field $\phi$ vanishes and 
it reduces to the four-dimensional Schwarzschild metric. 

Compared this solution with the four-dimensional three-parameter 
solution of $c_2 =1$ (see appendix A, where we give the 
three-parameter solution for an arbitrary dimensionality), 
we find that they are the same solution under the following 
identification  
\begin{align}
r_0^2 &= m^2 + q^2, \\
c_1^2 &= \frac{4q^2}{m^2 + q^2}. 
\label{q}
\end{align}
It is true that the three-parameter solution 
we have considered in this paper is a ten-dimensional one 
due to the consistency with the superstring theory,
but we can understand that the parameter $c_1$ relates to 
the scalar charge $q$ and this fact would be something helpful 
when we consider 
the physical meaning of $c_1$. 
Of course, this relation is not enough to clear the physical 
meaning of $c_1$, because this relation does not explain $c_1$ 
from the viewpoint of the source term of the three-parameter 
solution. That is, we need to know how $c_1$ is expressed 
in the Born-Infeld-type action, which makes it possible to relate $c_1$ to 
the boundary state. This issue will be also discussed in our 
next paper \cite{AKM2}.

\section*{Acknowledgments}
The authors would like to thank K. Ohta, 
H.~Kawai, T.~Tada, M.~Hayakawa, M. Sakagami and 
Y. Tsubo for helpful discussions. 
S.K. is also grateful to D. Ida 
for useful comments and advice. 
S.K. would also like to express his appreciation to 
Y. Himemoto for deep suggestions and continuous encouragement.  
This work of T.A. and S.M. is supported 
by Special Postdoctoral Researchers
Program at RIKEN.

\appendix 

\section{$D$-dimensional three-parameter solution}

In \cite{ZZ}, the $D$-dimensional three-parameter solution 
is given\footnote{
The $D$-dimensional four-parameter solution which possesses 
$ISO(p)\times SO(D-p-1)$ symmetry is also given in \cite{ZZ}.}. 
It is derived from the action, 
\begin{equation}
S = \frac{1}{2\kappa^2} \int d^{D} x \sqrt{-g} 
    \left[ 
    R -\frac{1}{2}(\del \phi)^2 
    -\frac{1}{2(p+2)!} e^{a\phi} |F_{p+2}|^2 
    \right], 
\label{3para-action}
\end{equation}
where $F_{(p+2)}$ denotes the $(p+2)$-form field strength 
and it relates to the $(p+1)$-form potential of RR-field 
$\mathcal{A}_{(p+1)}$ as $F_{(p+2)} = d \mathcal{A}_{(p+1)}$. 
The parameter $a$ is a dilaton coupling 
to the RR field strength.%
\footnote{
If we assume the gravity theory has stringy origin, 
the parameter $a$ is determined as 
$a=(D-2p-4)/(\sqrt{2(D-2)})$. 
}
  
The D\={D} system has the $ISO(1,p)\times SO(D-p-1)$ symmetry and 
the three-parameter solution also keeps it.  
The ansatz which follows this symmetry is  
\begin{align}
ds^2  = & \ e^{2A(r)} \eta_{\mu\nu}dx^{\mu}dx^{\nu} 
       +e^{2B(r)} \delta_{ij} dx^i dx^j \nonumber \\
      = & \ e^{2A(r)} \eta_{\mu\nu}dx^{\mu}dx^{\nu} 
      +e^{2B(r)} ( dr^2 + r^2 d\Omega_{(D-p-2)}^2 ), \\
\phi = &\ \phi(r), \\
\mathcal{A}^{(p+1)} 
         =& \ e^{\Lambda(r)} \ dx^0 \wedge dx^1 \wedge \cdots \wedge dx^p.
\label{3para-ansatz}
\end{align}
We represent the $D$-dimensional coordinates 
by $x^M$ $(M=0,1,\cdots,D-1)$, the $p$-brane world-volume coordinates  
$x^{\mu}$ $(\mu = 0,1,\cdots,p)$ and the transverse coordinates 
by $x^i$ $(i = p+1, \cdots, D-1)$ or, equivalently, by the polar 
coordinates $r, \theta_1, \cdots, \theta_{D-p-2}$ as we did in Sec.2.    
$r$ is defined as $r \equiv \sqrt{x^i x_i}$. 
The equations of motions of this system become
\begin{align}
A^{\prpr} + \left( (p+1)A^{\prime} 
+(D-p-3)B^{\prime} 
+\frac{D-p-2}{r}
\right) A^{\prime} 
&= \frac{D-p-3}{2(D-2)} S^2,  \\
B^{\prpr} + \left( (p+1)A^{\prime} 
+(D-p-3)B^{\prime} 
+\frac{2(D-p-3)+1}{r} 
\right)B^{\prime} 
+\frac{p+1}{r}A'
&= -\frac{p+1}{2(D-2)} S^2,  \\ 
(p+1)A^{\prpr} + (D-p-2)B^{\prpr} 
+(p+1)A^{\prime 2} + \frac{D-p-2}{r}B^{\prime} & \nonumber \\
-(p+1)A^{\prime} B^{\prime} 
+\frac{1}{2} \phi^{\prime 2} 
&= \frac{D-p-3}{2(D-2)} S^2,  \\
\phi^{\prpr} + \left(
 (p+1)A'+(D-p-3)B'+\frac{D-p-2}{r}
 \right) \phi^{\prime} 
&= -\frac{a}{2} S^2,  \\
 \Big( 
    \Lambda^{\prime} 
    e^{\Lambda + a\phi -(p+1)A +(D-p-3)B} r^{D-p-2}
 \Big)^{\prime} &= 0, 
\end{align}
where 
\begin{equation}
S^2 = \Lambda^{\prime 2} e^{2\Lambda + a\phi -2(p+1)A}. 
\end{equation}
The $D$-dimensional solution is given by 
\begin{align}
A(r) &= \frac{a c_1 (D-p-3)}{\Delta (D-2)} h(r) 
           -\frac{2(D-p-3)}{\Delta (D-2)} 
             \ln [\cosh(k h(r)) -c_2 \sinh (k h(r))], \\
B(r) &= \frac{1}{D-p-3}\ln[f_-(r) f_+(r)] \nonumber \\
         & \hspace{1cm}   -\frac{a c_1 (p+1)}{\Delta (D-2)} h(r) 
           +\frac{2(p+1)}{\Delta (D-2)} 
             \ln [\cosh(k h(r)) -c_2 \sinh (k h(r))], \\
\phi(r) &= \frac{2 c_1 (D-p-3)(p+1)}{\Delta (D-2)} h(r) 
              +\frac{2a}{\Delta} 
             \ln [\cosh(k h(r)) -c_2 \sinh (k h(r))], \\
e^{\Lambda(r)} &= -2\eta \left(\frac{c_2^2 -1}{\Delta} \right)^{1/2}
        \frac{\sinh(k h(r))}{\cosh(k h(r)) -c_2 \sinh (k h(r))}.
\end{align}
where 
\begin{align}
h(r) &\equiv \ln \left( \frac{f_-}{f_+}\right), \\
f_{\pm}(r) &\equiv 1 \pm \frac{r_0^{D-p-3}}{r^{D-p-3}}, \\
\Delta &\equiv \frac{2(D-p-3)(p+1)}{D-2} + a^2, \\
k &\equiv \pm \sqrt{\frac{\Delta}{4} 
                   \left[
                    \frac{2(D-p-2)}{D-p-3}
                     +\left( \frac{a^2}{\Delta} -1 \right)c_1^2
                   \right]
                 }, \\
\eta &= \pm 1.
\end{align}


\end{document}